\documentclass{article}
\newcommand{\bfr}{\begin{flushright}}
\newcommand{\efr}{\end{flushright}}
 
\begin{document}
\title{Global strings in five-dimensional supergravity
}
\author{ 
Miho Marui and
Kiyoshi Shiraishi\\
Department of Physics, Faculty of Science, Ochanomizu University,\\
1-1 Otsuka 2, Bunkyou-ku, Tokyo 112, Japan
}
\date{Physcs Letters {\bf B259} (1991) 58--62}
\maketitle
\begin{abstract}
We show the existence of solitonic solutions of five-dimensional
supergravity, which can be interpreted as global cosmic strings in our
universe. They possess the same mathematical structure as the stringy
cosmic strings studied by Greene, Shapere, Vafa and Yau, while the size
of the extra space and the value of the extra-space component of the
gauge field vary from place to place around the string in our model. We
also show that supersymmetry is partially broken in the presence of the
global strings.
\end{abstract}

Classical static solutions of higher-dimensional
theories including gravity have been studied extensively by many
authors during the past decade (see ref. \cite{1} on higher-dimensional
``black holes'', ref.~\cite{2} on monopoles, and ref.~\cite{3} on cosmic
strings). In the present paper, we examine some classical solutions
of the ${\cal N}=2$, $D=5$ supergravity theory \cite{4,5}, which
can be derived from ${\cal N}=1$, $D=11$ supergravity \cite{6},
by dimensional reduction.

Recently, Greene, Shapere, Vafa and Yau studied
\cite{7} vortex-like solutions in the system of multi-dimensional
gravity plus complex scalar fields, whose kinetic terms are
non-canonical. We can identify the scalar fields with the moduli of an
extra torus space. The scalar fields can also be interpreted as
combinations of dilaton and antisymmetric fields which are naturally
introduced when we consider string theories. They also discussed the
global structure of space in which propagation of strings is admitted.
Dabholkar et al.~\cite{8} and Strominger \cite{9} studied
another class of solitonic solutions in string theory
and discussed supersymmetry in the background of
the topological object.

``Strings'' in our five-dimensional model belong to
the same type as studied by Greene et al.~\cite{7}. It is
outstanding that the solution of moduli is given by
an arbitrary holomorphic function. A simple model
that they offered in their paper is a six-dimensional
model. We consider five-dimensional supergravity in
the present paper. One of the aims of this paper is to
provide the simplest, pedagogical model which realizes similar
solutions. Another aim is to discuss supersymmetry in the presence of
the string in the specified model. While the analysis is very similar
to refs.~\cite{8,9}, only holomorphicity of ``moduli'' is needed in
the present analysis.

We begin with the five-dimensional ${\cal N}= 2$ supergravity theory. The
supermultiplet consists of the f\"unfbein $e^A_M(x^N)$, the gravitino
$\psi^a_M(x^N)$ (where $a =1, 2$), and the gauge field $A_M(x^N)$. Our
notation is almost the same as that of ref.~\cite{5}. To make the
present paper self-contained, however, we shall exhibit the notational
convention.

One extra dimension is compactified on $S^1$, i.e.,
five-dimensional coordinates are separated as
$x^M=(x^\mu, x^5)=(x^\mu, \theta)$ with $0\le\theta<2\pi$. We use
$M, N,\dots (=\dot{1},\dots,\dot{5}), \mu, \nu,\dots
(=\dot{1},\dots,\dot{4})$, for world indices
and $A, B,\dots
(=1,\dots, 5), \alpha, \beta,\dots
(=1,\dots, 4)$
for Lorentz indices. 

The supersymmetric action is
\begin{eqnarray}
S^{(5)}&=&\int
d^5x\left(-\frac{1}{4}e\,e^M_Ae^N_BR^{AB}{}_{MN}-\frac{1}{4}eF_{MN}F^{MN}
\right.\nonumber \\
&
&-\frac{1}{6\sqrt{3}}\epsilon^{MNPSL}F_{MN}F_{PS}A_L
-\frac{1}{2}ie\bar{\psi}^a_M\Gamma^{MNP}D_N(\omega+\hat{\omega})\psi^a_P
\nonumber
\\ 
& &\left.-\frac{1}{32}i\sqrt{3}e(F_{MN}+\hat{F}_{MN})
\bar{\psi}^{Pa}(\gamma_
p\Gamma^{MN}\gamma_S-\gamma_S\Gamma^{MN}\gamma_P)\psi_a^S\right)\,.
\label{eq1}
\end{eqnarray}
The ``generalized Majorana spinor'' $\psi^a_M$, the covariant derivative
$D_N$, the totally antisymmetrized $\gamma$-matrices $\Gamma^{AB\cdots}$,
$\hat{F}_{MN}$, and  $\hat{\omega}_{MAB}$ are defined as follows: 
\begin{equation}
\bar{\psi}^a_M=\psi^a_MC\,,\quad
\bar{\psi}^1_M={\psi^2_M}^*\gamma^0\,,\quad
\bar{\psi}^2_M=-{\psi^1_M}^*\gamma^0\,,
\label{eq2a}
\end{equation}
where $C=i\gamma_1\gamma^5$ is the charge-conjugation matrix,
\begin{eqnarray}
& &D_N[\omega]\psi_{Pa}=\partial_N\psi_{Pa}+\frac{1}{4}\omega_{NPS}
\psi^S_a\,,\label{eq2b}\\
& &\Gamma^{A_1\cdots A_N}=\frac{1}{N!}\sum_{{\rm perm.}(A_1\cdots A_N)}
\mbox{sgn}(A_1\cdots A_N)\gamma^{A_1}\cdots \gamma^{A_N}\,,
\label{eq2c}\\
&
&\hat{F}_{MN}=F_{MN}+\frac{1}{2}i\sqrt{3}\bar{\psi}^a_M\psi_{Na}\,,
\label{eq2d}\\
&
&\hat{\omega}_{MAB}=\omega_{MAB}+\frac{1}{4}i\bar{\psi}^{Pa}\Gamma_{MABPS}
\psi^S_a\,.\label{eq2e}
\end{eqnarray}
To construct a string-like classical solution, we give
the following vacuum configurations of fields:
\begin{eqnarray}
& &\langle
e^\alpha_\mu\rangle=\frac{1}{\sqrt{b}}\bar{e}^\alpha_\mu(x^\mu)\,,
\quad\langle
e^5_{\dot{5}}\rangle=b(x^\mu)\,,\label{eq3a,b}\\
& &\langle
A_{\dot{5}}\rangle=A_{\dot{5}}(x^\mu)\,,\label{eq3c}\\
& &\langle\mbox{all other fields}\rangle=0\,, \label{eq3d}
\end{eqnarray}
where $b$ means the radius of $S^1$, the extra space.
These vacuum configurations are chosen to be consistent with the
equations of motion derived from the action (\ref{eq1}) in the tree
approximation. The configuration involving non-trivial dependence on
the spatial coordinates is very interesting to us. To analyze
such configurations, we look into the relevant part of
the four-dimensional action originating from the
Einstein-Maxwell system in five dimensions through
dimensional reduction:
\begin{equation}
S_{eff}^{(4)}=2\pi\int d^4x\frac{-\bar{e}}{4}
\left(\bar{R}^{(4)}-\frac{3}{2}\frac{\bar{\nabla}_\mu b\bar{\nabla}^\mu
b}{b^2}-2\frac{\bar{\nabla}_\mu A_{\dot{5}}\bar{\nabla}^\mu
A_{\dot{5}}}{b^2}\right)\,,\label{eq4}
\end{equation}
where letters with overbar indicate that those are defined in terms of
$\bar{e}^\alpha_\mu$ in eq.~(\ref{eq3a,b}). 

We define a complex scalar field (``moduli'') $\tau$, that is,
\begin{equation}
\tau=\tau_1+i\tau_2\,, \mbox{where }\tau_1=A_{\dot{5}} \mbox{ and } 
\tau_2=\frac{1}{2}\sqrt{3}b\,. \label{eq5}
\end{equation}
By the use of (\ref{eq5}), the effective action is written as
\begin{equation}
S_{eff}^{(4)}=2\pi\int d^4x\frac{-\bar{e}}{4}
\left(\bar{R}^{(4)}+\frac{3}{8}\frac{\bar{\nabla}_\mu\tau\bar{\nabla}^\mu
\tau}{(\tau-\bar{\tau})^2}\right)\,.\label{eq6}
\end{equation}
This action resembles one discussed by Greene et al.~\cite{7}, up to a
coefficient in the kinetic term of $\tau$. They considered torus
compactification in six-dimensional theory as a simple example. They
took an ansatz for the complex scalar field $\tau$ (moduli of the
two-torus)
\begin{equation}
\tau=\tau(x^2, x^3)\,, \label{eq7}
\end{equation}
and that for the metric of the four-dimensional theory
\begin{equation}
\bar{d}s^2=\bar{g}_{\mu\nu}dx^\mu dx^\nu
=(dx^1)^2-e^\phi[(dx^2)^2+(dx^3)^2]-(dx^4)^2\,,
\label{eq8}
\end{equation}
where $\phi=\phi(x^2, x^3)$. The metric of this form indicates that the
configuration is homogeneous along with the $x^4$-direction. Greene et
al.~solved the coupled Einstein equations in terms of the above variables
and then found a cosmic string solution dubbed as a ``stringy cosmic
string'' \cite{7}.

Now we adopt their assumptions (\ref{eq7}) and (\ref{eq8}) in
our model. The equation of motion for $\tau$ takes the
same form as theirs, i.e.,
\begin{equation}
\partial\bar{\partial}\tau+\frac{2\partial\tau\bar{\partial}
\bar{\tau}}{\tau-\bar{\tau}}=0\,, \label{eq9}
\end{equation}
while the non-trivial Einstein equation turns out to
be
\begin{equation}
\partial\bar{\partial}\phi=\frac{3}{(\tau-\bar{\tau})^2}
(\partial\tau\bar{\partial}
\bar{\tau}+\bar{\partial}\tau\partial
\bar{\tau})\,, \label{eq10}
\end{equation}
where $\partial\equiv\partial/\partial z$ and
$\bar{\partial}\equiv\partial/\partial \bar{z}$, for $z=x^2+ix^3$ and
$\bar{z}=x^2-ix^3$. 

Any holomorphic (or anti-holomorphic) function
$\tau$ is a solution to (\ref{eq9}), that is,
\begin{equation}
\bar{\partial}\tau=0\quad (\mbox{or } \partial\tau=0)\,. \label{eq11}
\end{equation}
We concentrate on the holomorphic solution of this
type in the present paper.
Substituting (\ref{eq11}) into (\ref{eq10}) then gives
\begin{equation}
\partial\bar{\partial}\phi=3\partial\bar{\partial}\ln\tau_2 \,.
\label{eq12}
\end{equation}
The solution to eq.~(\ref{eq12}) is therefore obtained as
\begin{equation}
\phi(z, \bar{z})=\ln \tau_2^3(z, \bar{z})+\ln f(z)+\ln f(\bar{z})\,,
\label{eq13}
\end{equation}
where $f(z)$ is some regular function.

From the observations so far, we can conclude that
the string-like solution, which is very similar to the
``stringy string'' obtained in ref.~\cite{7}, can be constructed in
terms of $\tau$ and $\phi$ in five-dimensional supergravity. The only
(mathematical) difference is the powers of $\tau$ in eq.~(\ref{eq13}),
which comes from the coefficient $3$ in eqs.~(\ref{eq10}) and
(\ref{eq12}).

We should note that $\tau+1$ is equivalent to $\tau$ because
the only local gauge equivalence permitted in the fifth
dimension is the identification $A_{\dot{5}}\sim A_{\dot{5}}+1$ in the
periodic dimension
\cite{10}.

As in ref.~\cite{7}, we can choose the Jacobi function
for $\tau$ in order to obtain a string solution with finite
energy density per unit length. In our model, we find
a discontinuity in $A_{\dot{5}}$ at the boundary of the fundamental
region,
$|\tau|= 1$. The appearance of singularities in the values of the fields
is a generic feature of vortices in non-linear sigma models \cite{11}.
Moreover the discontinuity in our model can be avoided if we
consider $S^1/Z_2$ as extra space, where the gauge fields
of opposite signs are identified with each other. We
leave the discussion on the property of an explicit solution written in
terms of elliptic functions for other occasions. In the rest of this
paper, we discuss a property of the string background without an
explicit functional form of the solution.

Let us discuss the supersymmetric structure of the
string-like configuration in five-dimensional supergravity.
Infinitesimal transformations of the supersymmetry on five-dimensional
fields are as follows
\cite{5}:
\begin{eqnarray}
\Delta e^A_M&=&-i\bar{\epsilon}\gamma^A\psi_M\,, \label{eq14a}\\
\Delta\psi_M&=&\partial_M\epsilon+\frac{1}{4}\hat{\omega}_{MAC}
\Gamma^{AC}\epsilon+\frac{1}{4\sqrt{3}}(\Gamma^{PQ}_M+4\gamma^P\delta^Q_M
)\hat{F}_{PQ}\epsilon\,,\label{eq14b}\\
\Delta A_M&=&-\frac{1}{2}i\sqrt{3}\bar{\epsilon}\psi_M\,, \label{eq14C}
\end{eqnarray}
where indices $a\, (=1, 2)$ of $\epsilon$ and $\psi$ are implicit.

In the dimensionally reduced theory, we wish to
concentrate our attention on zero modes, because a
matter of interest to us is the vacuum configuration
and the zero modes corresponding to unbroken symmetries. It is easy to
see that the transformation of bosonic fields in the string background
is zero, for the fermionic vacuum configuration is absent. On the
other hand, the transformation of ~ does not vanish
in general; thus we must investigate what form of
satisfies the equation $\langle \Delta\psi_M\rangle=0$, as in
refs.~\cite{8,9}.

Note that we can treat $\epsilon$ as a complex spinor instead of two
``generalized Majorana'' spinors. Hereafter we forget the label on
$\epsilon$.

By substituting eqs.~(\ref{eq3a,b}), (\ref{eq3c}),
(\ref{eq3d}), (\ref{eq7}) and (\ref{eq8}) into eq.~(\ref{eq14b}) and
using holomorphicity of $\tau$, i.e., the Cauchy-Riemann equation on
$\tau$ (eq.~(\ref{eq11})), we obtain
\begin{eqnarray}
& &\langle\Delta\psi_i\rangle=\langle\Delta\psi_{\dot{4}}\rangle
=\frac{e^{-\phi/2}}{2\sqrt{3}b}(\Gamma^{12}\partial_{\dot{3}}\tau_1
-\Gamma^{13}\partial_{\dot{2}}\tau_1)(1-\Gamma^{14})\epsilon=0\,,
\label{eq15a}\\
& &\langle\Delta\psi_{\dot{5}}\rangle
=\frac{\sqrt{b}e^{-\phi/2}}{\sqrt{3}}(\gamma^{3}\partial_{\dot{3}}\tau_1
+\gamma^{2}\partial_{\dot{2}}\tau_1)(1-\Gamma^{14})\epsilon=0\,,
\label{eq15b}\\
& &\langle\Delta\psi_z\rangle=\frac{1}{2}\left(
2\partial-\frac{1}{2}\partial\ln\frac{e^\phi}{\tau}(i\Gamma^{23})
+\frac{1}{2}\bar{\partial}\ln{\tau_2}\Gamma^{523}-{\partial}
\tau_2(i\gamma^5)
\right)\epsilon=0\,, \label{eq15c}\\
& &\langle\Delta\psi_{\bar{z}}\rangle=\frac{1}{2}\left(
2\bar{\partial}+\frac{1}{2}\bar{\partial}\ln\frac{e^\phi}{\tau}(i\Gamma^{23})
+\frac{1}{2}\bar{\partial}\ln{\tau_2}\Gamma^{523}+\bar{\partial}
\tau_2(i\gamma^5)
\right)\epsilon=0\,, \label{eq15d}
\end{eqnarray}
where
$\langle\Delta\psi_z\rangle\equiv\frac{1}{2}\langle
\Delta\psi_{\dot{2}}-i\Delta\psi_{\dot{3}}\rangle$
and $\langle\Delta\psi_{\bar{z}}\rangle\equiv\frac{1}{2}\langle
\Delta\psi_{\dot{2}}+i\Delta\psi_{\dot{3}}\rangle$.

We find, by inspection, solutions to (\ref{eq15a}), (\ref{eq15b}),
that is,
\begin{equation}
\Gamma^{14}\epsilon=\epsilon\,. \label{eq16}
\end{equation}
Furthermore, by the use of solution (\ref{eq13}) and the form
of (\ref{eq16}), it is proved that eqs.~(\ref{eq15c}), (\ref{eq15d}) lead
to the following form:
\begin{eqnarray}
\langle\Delta\psi_z\rangle&=&
\tau_2^{-1/4}\left(\frac{f}{\bar{f}}\right)^{\pm 1/4}
\partial\left[\tau_2^{1/4}\left(\frac{\bar{f}}{f}\right)^{\pm
1/4}\epsilon^{\pm}\right]=0\,, \label{eq17a}\\
\langle\Delta\psi_{\bar{z}}\rangle&=&
\tau_2^{-1/4}\left(\frac{f}{\bar{f}}\right)^{\pm 1/4}
\bar{\partial}\left[\tau_2^{1/4}\left(\frac{\bar{f}}{f}\right)^{\pm
1/4}\epsilon^{\pm}\right]=0\,, \label{eq17b}
\end{eqnarray}
where $f\equiv f(z)$, $\bar{f}\equiv f(\bar{z})$ and $\epsilon^{\pm}$
satisfy
$i\Gamma^{23}\epsilon^{\pm}=\pm\epsilon^{\pm}$. Therefore the
simultaneous solution to eqs.~(\ref{eq15a})--(\ref{eq15d}) is given by a
linear combination of $\hat{\epsilon}^+$ and $\hat{\epsilon}^-$,
\begin{equation}
\hat{\epsilon}^{\pm}=\tau_2^{-1/4}\left(\frac{f}{\bar{f}}\right)^{\pm
1/4}\epsilon_0^{\pm}\,,\label{eq18}
\end{equation}
where $\epsilon_0^{\pm}$ are constant spinors satisfying $(1-\Gamma^{14})
\epsilon_0^{\pm}=0$ and $(-1\pm i\Gamma^{23})\epsilon_0^{\pm}=0$.

Now we are led to the result that the background
given by the cosmic string in our model has partially
broken supersymmetry (i.e., supersymmetry associated with the restricted
form of $\epsilon$ (\ref{eq18}) remains unbroken). This conclusion is
independent of the explicit functional form of the solution, since we
have used only the holomorphicity of $\tau$ and the Einstein
equations.

Next, we see this symmetry breaking from the point
of view of the supersymmetry algebra. Generally
speaking, the extended supersymmetry algebra has a
central charge which is to give rise to partial symmetry breaking if
the background has non-trivial charge (see refs.~\cite{8,9} and
references therein). If we concentrate on string-like solutions, we
should define a supercharge per unit length and study the relation
between the central charge (per unit length) and supersymmetry breaking.

Since the vacuum configuration is translationally
invariant in the $x^4$ direction in the present model, we
consider the supercharge per unit length
\begin{equation}
Q(\epsilon')=\int_{\partial\Sigma}\bar{\epsilon}'\Gamma^{MNP}\psi_P
d\Sigma_{MN}\,, \label{eq19}
\end{equation}
where $\Sigma$ is a three-dimensional space-like surface,
which is given by the space spanned by coordinates
$(x^2, x^3, \theta)$ in our model. The supersymmetry transformation,
whose parameter is another spinor $\epsilon$, on $Q(\epsilon')$ is
\begin{equation}
\delta_\epsilon
Q(\epsilon')=\int_{\partial\Sigma}N^{MN}d\Sigma_{MN}=2
\int_{\Sigma}\nabla_MN^{MN}d\Sigma_N\,, \label{eq20}
\end{equation}
where
\begin{equation}
N^{MN}=\bar{\epsilon}'\Gamma^{MNP}\hat{D}_P\epsilon
\label{eq21}
\end{equation}
is Nester's form; here
\begin{equation}
\hat{D}_P=\partial_P+\frac{1}{4}\hat{\omega}_{PAC}\Gamma^{AC}+
\frac{1}{4\sqrt{3}}(\Gamma^{QR}_P+4\gamma^Q\delta^R_P)\hat{F}_{QR}
\label{eq22}
\end{equation}
is the supercovariant derivative.

In our case, the boundary $\partial\Sigma$ is considered as
$S^1\times S^1$, where the former means the spatial infinity
of the $x^2$-$x^3$ plane and the latter is the compactified
dimension. And then we can rewrite (\ref{eq20}) in terms of
contour integrals. Assuming that the spinors $\epsilon'$, $\epsilon$ are
constant at spatial infinity (say, $\epsilon'\rightarrow{\epsilon_0}'$, 
and $\epsilon\rightarrow{\epsilon_0}$ at spatial infinity) and the
radius of the compact space $b$ is independent of the azimuthal angle
$\varphi$ at spatial infinity $r\rightarrow\infty$ in the $x^2$-$x^3$
plane, we find
\begin{equation}
\int_{\partial\Sigma} N^{MN} d\Sigma_{MN}=
{\epsilon'}_0^+ (M-Z\Gamma^{14})\epsilon_0\,, 
\label{eq23}
\end{equation}
where $M$ and $Z$ can be interpreted as the mass per
unit length of the string and the central charge, respectively. Explicit
calculation reveals that $M$ and $Z$ take an identical value
\begin{equation}
M=Z=\sqrt{\tau_2}\frac{3}{2}\pi n\,, 
\label{eq24}
\end{equation}
where $n$ is the winding number of the string, defined
as
\begin{equation}
n\equiv\int_{\rm
spatial~infinity}\frac{\partial\tau_1}{\partial\varphi}d\varphi\,. 
\label{eq25}
\end{equation}

According to the expression (\ref{eq24}), both $M$ and $Z$
diverge. To describe the result in terms of finite
quantities, we must properly treat the length scale in
the $x^4$ direction, that is, the factor $b^{-1/2}$ which appeared in
(\ref{eq3a,b}).
In other words, we find a finite result referring to the effectively
four-dimensional ``barred'' metric system.

We observe here that $Z$ is proportional to
\begin{equation}
\int\tilde{F}_{ij}d\Sigma^{ij}
\label{eq26}
\end{equation}
(where $i,j$ run over $1, 2, 3, 5$ and $\tilde{~}$ means dual tensor),
which is known to arise as a part of the central charge for generic
${\cal N}\ge 2$ supergravity
\cite{12}. 

By following the ``standard'' procedure \cite{8,9}, we
can find
\begin{equation}
\delta_\epsilon Q(\epsilon)=\epsilon_0^+(M-Z\Gamma^{14})\epsilon_0
\ge 0\,.
\label{eq27}
\end{equation}
Since our solution has $M=Z$,
\begin{eqnarray}
& &\delta_\epsilon Q(\epsilon)>0\,,\quad \mbox{if }\Gamma^{14}
\epsilon_0=-\epsilon_0\,,
\label{eq28a}\\ & &\delta_\epsilon Q(\epsilon)=0\,,\quad \mbox{if
}\Gamma^{14}\epsilon_0=+\epsilon_0\,.
\label{eq28b}
\end{eqnarray}
This fact coincides with the previous analysis on the
transformation of fermionic fields; the above proves
that if the central charge $Z$ is non-zero our background admits
half-broken supersymmetry.

In future works we will clarify the symmetry among
the Kaluza-Klein excited modes, using a similar
technique as ref.~\cite{5}.

\bigskip

\bigskip

The authors would like to thank A.~Sugamoto for
reading this manuscript. K.S. is indebted to Soryuusi
shogakukai for financial support. He also would like
to acknowledge financial aid of Iwanami F\=ujukai.

\section*{Note added}
Recently we have been informed of a
string solution which involves non-zero torsion \cite{13}.
We thank Professor M. Hayashi for this information.


\end{document}